\documentstyle[aps]{revtex}

\begin{document}
\draft
\title{Generator Coordinate Method Calculations for Ground and First Excited
Collective States in $^{4}$He, $^{16}$O and $^{40}$Ca Nuclei}
\author{M.V. Ivanov, A.N. Antonov, M.K. Gaidarov}
\address{\it Institute of Nuclear Research and Nuclear Energy, Bulgarian\\
Academy of Sciences, Sofia 1784, Bulgaria}
\maketitle
\begin{abstract}
The main characteristics of the ground and, in particular, the first
excited monopole state in the ${}^{4}$He, ${}^{16}$O and ${}^{40}$Ca nuclei
are studied within the generator coordinate method using Skyrme-type
effective forces and three construction potentials, namely the
harmonic-oscillator, the square-well and Woods-Saxon potentials.
Calculations of density distributions, radii, nucleon momentum
distributions, natural orbitals, occupation numbers and depletions of the
Fermi sea, as well as of pair density and momentum distributions are
carried out. A comparison of these quantities for both ground and first
excited monopole states with the available empirical data and with the
results of other theoretical methods are given and discussed in detail.
\end{abstract}

\section{Introduction}
The study of nucleon-nucleon correlation effects is important part of the
contemporary nuclear physics \cite{AnH88,AnH93}. The basic idea of the
independent particle models (IPM) consists in the assumption that nucleons
move independently in a mean field created by the same nucleons.
Consequently, each coherent motion of the nucleons creates changes in the
mean field. The Hartree-Fock approximation accounts only partially for the
dynamic nucleon correlations. There exist many experimental data (see e.g.
the review in \cite{AnH93}) showing that the IPM are unable to describe
basic nuclear characteristics. For instance, the nucleon momentum and
density distributions in nuclei cannot be reproduced simultaneously
\cite{Boh80,Jam85}. This is also the case with the occupation numbers, the
hole-state spectral functions, with characteristics of nuclear reactions
and others \cite{AnH88,AnH93,Fro87,Lap91,Lap92}. This imposes the
development of correlation methods going beyond the limits of the
mean-field approximation (MFA) which account for nucleon-nucleon
correlations (see e.g. \cite{AnH93}). This can be reached by extending the
class of the trial functions which is used at the diagonalization of the
nuclear Hamiltonian.

The generator coordinate method (GCM) \cite{Hil53,Gri57} is one of the
methods beyond the MFA which have been applied successfully to studies of
the collective nuclear motions. For instance, the GCM has been applied to
investigate the breathing-mode giant monopole resonance within the
framework of the relativistic mean-field theory in \cite{Sto94}. In it the
constrained incompressibility and the excitation energy of isoscalar giant
monopole states were obtained for finite nuclei with various sets of
Lagrangian parameters. An extension of the method of Ref. \cite{Sto94} by
using a more general ansatz for the generating functions of the GCM and by
including the isovector giant monopole states has been done in
\cite{Vre96}. The use of the Skyrme effective forces has simplified the
study of the monopole, dipole and quadrupole isoscalar and isovector
vibrations in light double magic nuclei \cite{Flo75,Flo76}. An approach to
the GCM using square-well and harmonic-oscillator construction potentials
has been applied to calculate the energies of the ground and first monopole
excited state, as well as the density and the nucleon momentum
distributions in the ground state of ${}^{4}$He, ${}^{16}$O and ${}^{40}$Ca
nuclei \cite{AnC86,AnB88}. In \cite{AnB90} the occupation numbers, the
depletion of the Fermi sea and the natural orbitals (NO) which diagonalize
the ground state one-body density matrix (ODM) have been calculated. The NO
related to the ground state and the single-particle potentials
corresponding to them have been studied in detail in \cite{AnK94}. In
\cite{AnH93,AnB91} the two-nucleon center-of-mass and relative motion
momentum distributions of $n-p$ pairs in the ${}^{4}$He, ${}^{16}$O and
${}^{40}$Ca nuclei have been calculated. The existence of high-momentum
components in the one-nucleon and two-nucleon momentum distributions in the
case of the square-well construction potential with infinite walls within
the GCM has been obtained. The studies of the energies, the nucleon
momentum and density distributions in ${}^{4}$He and ${}^{16}$O nuclei have
been extended by means of a two generator coordinate scheme in
\cite{AnS94}.

In the last years giant resonances with various multipolarities different
from the well-known dipole resonance have been discovered. Among these
collective excitations the monopole excitations take a particular place.
The isoscalar giant monopole resonances or the so-called breathing
vibrational states (with ${I}^{\pi}={0}^{+}$ at energies of approximately
13 to 20 MeV) have been well established experimentally (e.g.
\cite{You81,Spe81,Voi85,Lui85,LuB86,Bra87,Sha88,Bor89}) and have been
considered to be compressional nuclear vibrations. Their study concerns the
important problem of the compressibility of finite nuclei and nuclear
matter (e.g. \cite{AnN91,Bru70,Pan70,Tre81,VoE85,Sha89}). The description
of such states is related mainly to general characteristics of the nuclei
and weakly to the peculiarities of nuclear structure. Recently, in a series
of articles Bishop et al.(see \cite{Bis93} and references therein) applied
the translationally invariant cluster (TIC) method to light nuclei, in
particular to the ${}^{4}$He nucleus. The basic properties, such as the
energies and the density distributions of the ground and first excited
breathing mode state in ${}^{4}$He have been considered.

The ${}^{4}$He nucleus has a well-established spectrum of excited states.
Calculations for the monopole oscillations of helium which practically
involve the whole nuclear volume have been performed on the base of the
nonlinear time-dependent Hartree-Fock ~(TDHF) approximation
 \cite{Wu 97}. Its small amplitude limit, namely the random phase
approximation ~(RPA), has been used extensively to describe nuclear
collective motion \cite{Col92}. The TDHF method \cite{Wu 99} and the
relativistic RPA \cite{Ma 97} have been recently used to extend the study
of isoscalar monopole modes in finite nuclei up to ${}^{208}$Pb. It has
been found that some effective Lagrangians can describe ground states and
giant resonances as well, and in particular, they can predict correctly
breathing mode energies in medium and heavy nuclei. In general, it has been
pointed out that when going from heavy to lighter systems the trend is that
the collectivity becomes weaker. This fact poses the question about the
role of different kinds of nucleon-nucleon correlations corresponding to
single-particle and collective motion modes and, therefore, the
investigations on the correlation effects became an important task in the
nuclear theory.

The aim of the present work is to study the main characteristics of the
ground and, in particular, of the first excited monopole state in the
${}^{4}$He, ${}^{16}$O and ${}^{40}$Ca nuclei within the GCM using
Skyrme-type effective forces. It concerns the energies, the density
distributions and radii, the nucleon momentum distributions, the natural
orbitals and occupation numbers, as well as the pair center-of-mass and
relative density and momentum distributions. As known, the natural orbital
representation (NOR)\cite{Lwd55} gives a model-independent effective
single-particle picture for the correlated states. We emphasize that in our
work this is done also for the first excited monopole state in the nuclei
considered on the basis of the ODM calculated for this state. It is known
that the results of the GCM calculations depend on the type of the
construction potential used to define the generating function in the
method. In the present work we use three construction potentials, namely
the harmonic oscillator ((HO) where the oscillator parameter is a generator
coordinate), the square-well with infinite walls ((SW) where the radius of
the well is a generator coordinate) and Woods-Saxon ((WS1) where the
diffuseness of the Woods-Saxon well is a generator coordinate and (WS2)
where the radial parameter of Woods-Saxon well is a generator coordinate).
The values of the radial parameter in the WS1 case and the diffuseness
parameter in the WS2 case are taken from \cite{BH 91}. In our opinion, the
calculations performed in this work can give an essential information about
beyond MFA effects (accounted for in the considered approach to the GCM) on
the mentioned characteristics. The results are compared with the available
empirical data and with the calculations of other theoretical methods.

The basic relations of the GCM are given in Section 2. The results of the
calculations and the discussion are presented in Section 3. The conclusions
from the work are given in Section 4.

\section{Basic Relations in GCM}
In the case of one real generator coordinate $x$ the trial many-body wave
function in the GCM is written as a linear combination\cite{Gri57}:
\begin{equation}
{\Psi}({\bf{r}}_i)=
\int_{0}^{\infty}{f(x){\Phi}(\{{\bf{r}}_i\},x)dx}\,,\;\;i=1..A\,,
\end{equation}
where ${\Phi}(\{{\bf{r}}_i\},x)$ is the generating function, $f(x)$ is the
generator or weight function and $A$ is the mass number of the nucleus.

The application of the Ritz variational principle ${\delta}E=0$ leads to
the Hill-Wheeler integral equation for the weight function:
\begin{equation}
\int_{0}^{\infty}{[{\cal H}(x,x')-EI(x,x')]f(x')dx'}=0\,,
\end{equation}
where
\begin{equation}
{\cal H} (x,x')=<{\Phi}(\{{\bf{r}}_i\}\,x)|\hat{H}|{\Phi}
(\{{\bf{r}}_i\}\,x')>\,,
\end{equation}
and
\begin{equation}
I(x,x')=<{\Phi}(\{{\bf{r}}_i\}\,x)|{\Phi}(\{{\bf{r}}_i\}\,x')>\,
\end{equation}
are the energy and overlap kernels, respectively, and $\hat{H}$ is the
Hamiltonian of the system.

If the generating function ${\Phi}(\{{\bf{r}}_i\}\,x)$ for a $N=Z$ nucleus
is a Slater determinant built up from single-particle wave functions
${\varphi}_{\lambda}(\bf{r},x)$ corresponding to a given construction
potential then the energy kernel (3) has the form\cite{Bri66}:
\begin{equation}
{\cal H}(x,x')=<{\Phi}(\{{\bf{r}}_i\}\,x)|{\Phi}(\{{\bf{r}}_i\}\,x')>
\int_{}^{}{H(x,x',{\bf{r}})}d{\bf{r}}\,.
\end{equation}
In the case of the Skyrme-like forces (for nucleus with $Z=N$ and without
Coulomb and spin-orbital interaction) $H(x,x',\bf{r})$ is given by:
\begin{equation}
H(x,x',{\bf{r}})=\frac{{\hbar}^2}{2m}T+
\frac{3}{8}{t_0}{{\rho}^2}+
\frac{1}{16}(3{t}_1+5{t}_2)({\rho}T+{\bf{j}}^2)+
\frac{1}{64}(9{t}_1-5{t}_2){({\nabla}{\rho})}^2+
\frac{1}{16}{t}_3{\rho}^{2+{\sigma}}\,,
\end{equation}
where ${t}_0,{t}_1,{t}_2,{t}_3,{\sigma}$ are the Skyrme-force parameters.
The density ${\rho}$, the kinetic energy $T$ and the current density
${\bf{j}}$ are defined by:
\begin{equation}
{\rho}(x,x',{\bf{r}})=4\sum_{{\lambda},{\mu}=1}^{A/4}
{{{(N^{-1})}_{{\mu}{\lambda}}}{{\varphi}_{\lambda}^{*}}({\bf{r}},x)
{{\varphi}_{\mu}}({\bf{r}},x')}\,,
\end{equation}
\begin{equation}
T(x,x',{\bf{r}})=4\sum_{{\lambda},{\mu}=1}^{A/4}
{{{(N^{-1})}_{{\mu}{\lambda}}}{{\nabla\varphi}_{\lambda}^{*}}({\bf{r}},x)\cdot
{{\nabla\varphi}_{\mu}}({\bf{r}},x')}\,,
\end{equation}
\begin{equation}
{\bf{j}}(x,x',{\bf{r}})=2\sum_{{\lambda},{\mu}=1}^{A/4}
{{{(N^{-1})}_{{\mu}{\lambda}}}\{{{\varphi}_{\lambda}^{*}}({\bf{r}},x)
{{\nabla\varphi}_{\mu}}({\bf{r}},x')-[{\nabla\varphi}_{\lambda}^{*}({\bf{r}},x)]
{{\varphi}_{\mu}}({\bf{r}},x')\}}\,,
\end{equation}
where
\begin{equation}
{N}_{{\lambda}{\mu}}(x,x')=
\int_{}^{}{{\varphi}_{\lambda}^{*}({\bf{r}},x){\varphi}_{\mu}({\bf{r}},x')}
d{\bf{r}}\,.
\end{equation}
The overlap kernel (4) is given by:
\begin{equation}
I(x,x')={[det({N}_{{\lambda}{\mu}})]}^4\,.
\end{equation}
Solving the Hill-Wheeler equation(2) one can obtain the solutions
${f}_0,{f}_1,...$ for the weight functions which correspond to the
eigenvalues of the energy ${E}_0,{E}_1...$.

The one-body density matrix ${\rho}_{i}({\bf{r}},{\bf{r'}})$ of the ground
($i=0$) and the first excited monopole ($i=1$) states in this GCM scheme
has the form:
\begin{equation}
{\rho}_{i}({\bf{r}},{{\bf{r'}}})=\int\!\!\int
{{f}_{i}(x){f}_{i}(x')I(x,x'){\rho}(x,x',{\bf{r}},{\bf{r'}})dxdx'}
\,,\;\;i=0,1\,,
\end{equation}
where
\begin{equation}
{\rho}(x,x',{\bf{r}},{{\bf{r'}}})=4\sum_{{\lambda},{\mu}=1}^{A/4}
{{{(N^{-1})}_{{\mu}{\lambda}}}{{\varphi}_{\lambda}^{*}}({\bf{r}},x)
{{\varphi}_{\mu}}({{\bf{r'}}},x')}\,.
\end{equation}
It follows from (12) that the nucleon density distribution
${\rho}_{i}({\bf{r}})$ and the nucleon momentum distribution
${n}_{i}(\bf{k})$ of the ground and the first excited monopole states can
be expressed as:
\begin{equation}
{\rho}_{i}({\bf{r}})=\int\!\!\int
{{f}_{i}(x){f}_{i}(x')I(x,x'){\rho}(x,x',{\bf{r}})dxdx'}
\,,\;\;i=0,1
\end{equation}
and
\begin{equation}
{n}_{i}({\bf{k}})=\int\!\!\int
{{f}_{i}(x){f}_{i}(x')I(x,x'){\rho}(x,x',{{\bf{k}}})dxdx'}
\,,\;\;i=0,1\,,
\end{equation}
where
\begin{equation}
{\rho}(x,x',{\bf{r}})={\rho}(x,x',{\bf{r}},{{{\bf{r'}}}}={\bf{r}})\,,
\end{equation}
\begin{equation}
{\rho}(x,x',{{\bf{k}}})=4\sum_{{\lambda},{\mu}=1}^{A/4}
{{{(N^{-1})}_{{\mu}{\lambda}}}{{\tilde{\varphi}}_{\lambda}^{*}}({{\bf{k}}},x)
{{\tilde{\varphi}}_{\mu}}({{\bf{k}}},x')}\,
\end{equation}
and ${\tilde{\varphi}}({{\bf{k}}},x)$ is the Fourier transform of
${\varphi}({\bf{r}},x)$.

The rms radii can be calculated using the expression:
\begin{equation}
{r}_{rms}^{(i)}=\sqrt{\frac{\int{{r}^{4}{\rho}_{i}(r)dr}}
{\int{{r}^{2}{\rho}_{i}(r)dr}}}
\,,\;\;i=0,1\,.
\end{equation}

As shown e.g. in \cite{AnH88,AnC89} the attractive properties of the
single-particle description can be preserved in the correlation methods
using the natural orbital representation \cite{Lwd55}. In it the one-body
density matrix has the simple form:
\begin{equation}\label{den}
{\rho}_{i}({\bf{r}},{{{\bf{r'}}}})=\sum_{\alpha}^{}{{n}_{\alpha}^{(i)}
{\psi}_{\alpha}^{(i)*}({\bf{r}}){\psi}_{\alpha}^{(i)}({{{\bf{r'}}}})}
\,,\;\;i=0,1\,,
\end{equation}
where the natural orbitals (NO) ${\psi}_{\alpha}^{(i)}({\bf{r}})$ form a
complete orthonormal set of single-particle wave functions which
diagonalize the density matrix ${\rho}_{i}({\bf{r}},{{{\bf{r'}}}})$ for the
ground and the first excited monopole states. The natural occupation
numbers (ON) ${n}_{\alpha}^{(i)}$ for the state ${\alpha}$ satisfy the
conditions:
\begin{equation}
0{\leq}{n}_{\alpha}^{(i)}{\leq}1,\;\;\;\;\;
\sum_{\alpha}^{}{{n}_{\alpha}^{(i)}}=A\,.
\end{equation}
It is seen from (\ref{den}) that the natural orbitals and the occupation
numbers can be found solving the equation:
\begin{equation}
\int{{\rho}_{i}({\bf{r}},{{{\bf{r'}}}}){\psi}_{\alpha}^{(i)}({{{\bf{r'}}}})d{{{\bf{r'}}}}}=
{n}_{\alpha}^{(i)}{\psi}_{\alpha}^{(i)}({\bf{r}})
\,,\;\;i=0,1\,.
\end{equation}

For nuclei with total spin $J=0$ the one-body density matrix can be
diagonalized in the $\{l\,j\,m\}$ subspace ($l,j,m$ being the quantum
numbers corresponding to the angular momentum, total momentum and its
projection). In the case of nuclei with spherical symmetry the natural
orbitals have the form:
\begin{equation}
{\psi}_{nlm}^{(i)}({\bf{r}})=
{R}_{nl}^{(i)}(r){Y}_{lm}({\theta},{\varphi})=
\frac{{u}_{nl}^{(i)}(r)}{r}{Y}_{lm}({\theta},{\varphi})
\,,\;\;i=0,1\,,
\end{equation}
where ${R}_{nl}^{(i)}(r)$ is the radial part of the NO's and
${Y}_{lm}({\theta},{\varphi})$ is the spherical function. The substitution
of ${\psi}_{nlm}^{(i)}({\bf{r}})$ from (22) and
${\rho}_{i}({\bf{r}},{{{\bf{r'}}}})$ from (12) in equation (21) and the
integration over the angular variables give the following equation for the
radial part ${u}_{nl}^{(i)}(r)$ of the NO's and the occupation numbers
${n}_{nl}^{(i)}$:
\begin{equation}
\int_{0}^{\infty}{{K}_{l}^{(i)}({\bf{r}},{{{\bf{r'}}}}){u}_{nl}^{(i)}(r')dr'}=
{n}_{nl}^{(i)}{u}_{nl}^{(i)}(r)
\,,\;\;i=0,1\,,
\end{equation}
where
\begin{equation}
{K}_{l}^{(i)}({\bf{r}},{{{\bf{r'}}}})=rr'\int\!\!\int
{f}_{i}(x){f}_{i}(x')I(x,x')\sum_{n,n'}
{{(N^{-1})}_{nl,n'l}}{{\cal R}}_{n'l}^{*}(r,x)
{{\cal R}}_{nl}(r',x')dxdx'
\,,\;\;i=0,1\,.
\end{equation}
The summation in (24) is performed over all single-particle wave functions
with given $l$ forming the Slater determinant of the generating function
and ${{\cal R}}_{nl}(r,x)$ are the radial parts of these functions.

The nucleon density distribution ${\rho}_{i}(r)$ and the nucleon momentum
distribution ${n}_{i}(k)$ for nuclei with $J=0$ can be written in the
following form in the NOR:
\begin{equation}
{\rho}_{i}(r)=\frac{1}{4{\pi}}\sum_{l}2(2l+1)
\sum_{n}{n}_{nl}^{(i)}{|{R}_{nl}^{(i)}(r)|}^2
\,,\;\;i=0,1\,,
\end{equation}
\begin{equation}
{n}_{i}(k)=\frac{1}{4{\pi}}\sum_{l}2(2l+1)
\sum_{n}{n}_{nl}^{(i)}{|{\tilde{R}}_{nl}^{(i)}(k)|}^2
\,,\;\;i=0,1\,,
\end{equation}
where
\begin{equation}
{\tilde{R}}_{nl}^{(i)}(k)={(\frac{2}{\pi})}^{1/2}{(-i)}^{l}
\int_{0}^{\infty}{{r}^{2}{j}_{l}(kr){R}_{nl}^{(i)}(r)dr}
\,,\;\;i=0,1
\end{equation}
is the radial part of the NO in the momentum space and ${j}_{l}(kr)$ are
spherical Bessel functions of order $l$.

The depletion of the Fermi sea can be defined by the expression:
\begin{equation}
{{\cal D}}^{(i)}=\frac{4}{A}\sum_{j,\alpha\in\{F\}}
(2j+1)(1-{n}_{\alpha}^{(i)})
\,,\;\;i=0,1\,,
\end{equation}
where $\{F\}$ refers to the Fermi sea.

The rms radius of the NO ${\psi}_{\alpha}^{(i)}({\bf{r}})$ is given by the
expression:
\begin{equation}
<{r}_{\alpha }^{(i)}>=\sqrt {\frac
{\int{{r}^{2}{|{\psi }_{\alpha }^{(i)}({\bf{r}})|}^2d{\bf{r}}}}
{\int{{|{\psi }_{\alpha }^{(i)}({\bf{r}})|}^2d{\bf{r}}}}}
\, , \;\; i=0,1\,.
\end{equation}

The two-body density matrix $\rho ^{(2)}(\xi _{1},\xi _{2};\xi _{1}^{\prime
},\xi _{2}^{\prime })$ in the GCM has the form:
\begin{equation}
\rho ^{(2)}(\xi _{1},\xi _{2};\xi _{1}^{\prime },\xi _{2}^{\prime })=\int
dxf^{\ast }(x)\int dx^{\prime }f(x^{\prime })\rho ^{(2)}(x,x^{\prime };\xi
_{1},\xi _{2};\xi _{1}^{\prime },\xi _{2}^{\prime })\ ,\label{tbdm1}
\end{equation}
where the coordinate $\xi $ includes the spatial coordinate ${{{{\bf {r}}}%
}}$, as well as the spin ($s$) and isospin ($\tau $) variables. If the
generating wave function is a Slater determinant constructed from a
complete set of one-particle wave functions ${{{\varphi }_{i}}(x,\xi )}$
the matrix $\rho ^{(2)}(x,x^{\prime };\xi _{1},\xi _{2};\xi _{1}^{\prime
},\xi
_{2}^{\prime })$ can be expressed as \cite{Lwd55}:\ \
\begin{equation}
\rho ^{(2)}(x,x^{\prime };\xi _{1},\xi _{2};\xi _{1}^{\prime },\xi
_{2}^{\prime })=\frac{I(x,x^{\prime })}{2}\det \left(
\begin{array}{cc}
\rho ^{(2)}(x,x^{\prime };\xi _{1},\xi _{1}^{\prime }) & \rho
^{(2)}(x,x^{\prime };\xi _{1},\xi _{2}^{\prime }) \\
\rho ^{(2)}(x,x^{\prime };\xi _{2},\xi _{1}^{\prime }) & \rho
^{(2)}(x,x^{\prime };\xi _{2},\xi _{2}^{\prime })
\end{array}
\right) \ ,
\end{equation}
where
\begin{equation}
\rho ^{(2)}(x,x^{\prime };\xi ,\xi ^{\prime })=%
\mathrel{\mathop{\stackrel{A}{\sum }}\limits_{k,l=1}}%
\left( N^{-1}\right) _{lk}{{{\varphi }_{k}^{\ast }}(x,\xi ){{\varphi }_{l}}%
(x^{\prime },\xi ^{\prime })}\ ,
\end{equation}
$\left( N^{-1}\right) _{lk}$ is the inverse matrix of:
\begin{equation}
N_{kl}(x,x^{\prime })=%
\mathrel{\mathop{\sum }\limits_{s,\tau }}%
\int d{{{\bf {r}}{\varphi }_{k}^{\ast }}(x,\xi ){{\varphi }_{l}}(x^{\prime
},\xi )}\ ,
\end{equation}
and
\begin{equation}
I(x,x^{\prime })=\det \left( N_{kl}\right) \ .
\end{equation}
The two-particle emission experiments require some knowledge of physical
quantities associated with the TDM. For example, using the two-body density
matrix $\rho ^{(2)}$ in coordinate space, Eq.(\ref{tbdm1}), one can define
the pair center-of-mass local density distribution:
\begin{equation}
{\bf \;\;}\rho ^{(2)}({\bf R)=}\int \rho ^{(2)}({\bf {R+s/}}2{\bf ,{R-s/}}2%
{\bf )}d{\bf s\ } \label{denrelR}
\end{equation}
and the pair relative local density distribution:
\begin{equation}
\rho ^{(2)}({\bf s) =}\int \rho ^{(2)}({\bf R+s/}2{\bf ,R-s/}2{\bf )}d{\bf %
R\,.} \label{denrelr}
\end{equation}
In momentum space the associated pair center-of-mass and relative momentum
distributions can be defined:
\begin{equation}
n^{(2)}({\bf K)=}\int n^{(2)}({\bf K/}2{\bf +k,K/}2{\bf -k)}d{\bf
k\,,}\label{denrelK}
\end{equation}
\begin{equation}
n^{(2)}({\bf k)=}\int n^{(2)}({\bf K/}2{\bf +k,K/}2{\bf -k)}d{\bf K\,.}
\label{denrelk}
\end{equation}

The physical meaning of $\rho ^{(2)}({\bf s)}$ and $n^{(2)}({\bf k)}$ is
the probability to find two particles displaced of a certain relative
distance ${\bf s=r}_{1}{\bf -r}_{2}$ or moving with relative momentum ${\bf
k=(k}_{1}{\bf -k}_{2})/2$, respectively, while $\rho ^{(2)}({\bf R)}$ and
$n^{(2)}({\bf K)}$ represent the probability to find a pair of particles
with center-of-mass coordinate ${\bf R=}\left( {\bf r}_{1}+{\bf
r}_{2}\right) {\bf /}2$ or center-of-mass momentum ${\bf K=k}_{1}{\bf
+k}_{2}$.

\section{Results and Discussion}
\subsection{The ground and first excited collective state energies}
The Hill-Wheeler equation (2) is solved using a discretization procedure
similar to that of Refs. \cite{Flo75,Flo76}. The values of the Skyrme-force
parameters in (6) in the case of square-well construction potential are the
same as in \cite{AnB88}. They are determined to give an optimal fit to the
binding energies of $^{4}$He, $^{16}$O and $^{40}$Ca obtained from the
Hill-Wheeler equation (2). The parameter set values $t_{0}=-2765,
t_{1}=383.94, t_{2}=-38.04, t_{3}=15865, {\sigma}=1/6$ lead to the energies
of the ground $E_{0}$ and the first monopole excited $E_{1}$ states
(without the Coulomb energy) shown in Table \ref{t1}. In the case of the
harmonic-oscillator and Woods-Saxon construction potentials SkM* parameter
set values ($t_{0}=-2645, t_{1}=410, t_{2}=-135, t_{3}= 15595,
{\sigma}=1/6$)\cite{Bar82} giving realistic binding energies are used. It
can be seen from Table \ref{t1} that the energies obtained with WS1 are
close to that obtained with HO, while the energies corresponding to WS2 are
closer to that calculated with SW construction potential.

The values of the excitation energies ${\Delta}E=E_{1}-E_{0}$ of the first
monopole state (${I}^{\pi}={0}^{+}$) calculated within the GCM are given
and compared with other calculations and some experimental data in Table
\ref{t2}. It is seen that the GCM results with HO construction potential
and in the WS1 case are in good agreement with the result of Brink and Nash
\cite{Bri63} for $^{16}$O nucleus and they are closer to the experimental
values than the GCM results with the SW construction potential. The energy
of the first excited $0^{+}$ in $^{4}$He obtained in the present work is
compared in Table \ref{t2} with the recent refined calculations of Bishop
et al \cite{Bis93} within the TIC method as well as with the results of the
coupled rearrangement of the channels method \cite{Kam89}.

\subsection{The one-body density matrix ${\rho}({\bf{r}},{{{\bf{r'}}}})$ of
the ground and first excited monopole states}
\subsubsection{The nucleon density distribution ${\rho}({{\bf{r}}})$}
The one-body distribution function corresponding to the density
distribution ${\rho}_{i}({\bf{r}})$ from (14) is given by the expression:
\begin{equation}\label{pro}
g_{i}(r)=4{\pi}{r^2}{\rho _{i}(r)}
\, ,\; \; i=0,1 \; .
\end{equation}
The function ${g}_{i}(r)$ of the ground (i=0) and the first excited
monopole (i=1) states of $^{4}$He calculated in the GCM are compared in
Figure 1 with those obtained in the TIC method. It can be seen that there
is a large difference in the behaviour of the function $g(r)$ for both
states. This is due to the strong increase of the size of the nucleus, as
is shown in Table \ref{t3}: the rms radius increases from 1.89 fm for the
ground state to 3.00 fm for the first excited monopole state in the case of
the HO, from 1.77 fm to 2.86 fm in the case of the SW and from 1.85 fm to
3.29 fm when using WS2. The significantly larger values of rms radii for
the first excited state in $^{4}He$ nucleus in respect to their ground
state values support the breathing-mode interpretation. At the same time,
the nature of the first excited monopole state in $^{4}He$ is not still
well understood. An attempt to solve this problem is made in Ref.
\cite{Cso97} where it is concluded that the $i=1$ state is dominated by the
$^{3}H+p$ configuration. The rms radii of the ground and of the first
excited monopole states in $^{16}$O and $^{40}$Ca are also presented in
Table \ref{t3}. We note that the rms radii of the ground state in these
nuclei obtained with Woods-Saxon construction potential are very close to
the experimental values.

For the first excited monopole state the one-body distribution function
$g_{1}(r)$ calculated within the GCM using the WS2 potential and that one
calculated within the TIC method are in good agreement. In contrast with
this state, for a ground state both methods yield functions $g_{0}(r)$
which differ significantly. This is due to the spurious centre-of-mass
motion correction for $^{4}$He which is accounted for in the TIC method
\cite{Bis93}. At the same time effect doesn't influence substantially the
first excited monopole-state density.

The nucleon density distributions of $^{16}$O and $^{40}$Ca of the ground
${\rho}_{0}$ and the first excited monopole ${\rho}_{1}$ states obtained
within the GCM are presented in Figures 2 and 3, respectively. For the
first excited monopole state the density distribution ${\rho}_{1}(r)$
decreases in the central region of both nuclei and increases for large $r$.
This behaviour corresponds to the increase of the rms radius in the first
excited state. It can be seen also that the difference in the rms radii
${\Delta}{r}_{rms}={r}_{rms}^{(1)}-{r}_{rms}^{(0)}$ decreases with the
increase of the mass number.

\subsubsection{The nucleon momentum distribution ${n}({{\bf{k}}})$}
The nucleon momentum distribution ${n}({{\bf{k}}})$ is one of the nuclear
quantities which are sensitive to various types of nucleon-nucleon
correlations. The momentum distributions for the ground and first excited
monopole states of $^{16}$O have been calculated in the GCM using Eqs.(15)
and (17) and are shown in Figure 4. As can be seen there is no essential
difference between the behaviour of the ${n}({{\bf{k}}})$ in the ground and
in the first excited monopole states.

It is seen from Figure 4 that the nucleon momentum distribution
${n}({{\bf{k}}})$ obtained in the GCM depends on the construction
potentials considered. The use of SW potential leads to an existence of
high-momentum tail of ${n}({{\bf{k}}})$ which is not the case when the HO
and WS potentials are used. Here we would like to note that:

(i) the single-particle wave functions corresponding to the SW potential
contain themselves high-momentum components due to the particular form of
the potential and they are reflected in the GCM result,

(ii) the lack of substantial high-momentum tail in ${n}({{\bf{k}}})$ when
using HO and WS construction potential leads to the conclusion that the
nucleon-nucleon correlations included in the GCM approach are not of
short-range type. An additional study of the correlations accounted for in
the GCM concerns some two-body characteristics and we will discuss the
results for them in subsection C.

\subsubsection{The natural orbitals and occupation numbers}
The natural orbitals and occupation numbers have been calculated using
Eqs.(21)-(24). A discretization procedure with respect to both $r$ and $r'$
has been applied solving Eq.(23) and the resulting matrix eigenvalue
problem has been solved numerically. The natural orbitals obtained in the
coordinate space are plotted in Figure 5 for the $^{16}$O nucleus. The
corresponding rms radii and the natural occupation numbers of $^{4}$He,
$^{16}$O and $^{40}$Ca are shown in Tables 4 and 5, respectively.

One can see from Table \ref{t4} and from the shape of the NO's given in
Figure 5 that the rms radii of the natural hole-states increase from the
ground state to the first excited state. The occupation numbers of the
natural hole-states decrease in the first excited monopole state, while
those corresponding to the natural particle-states increase (see Table
\ref{t5}). As can be seen from the same Table \ref{t5}, the depletion of
the Fermi sea increases in the first excited state. This is strongly
expressed for the $^{4}$He nucleus. The calculated depletion decreases with
the increase of the mass number.

\subsection{The pair center-of-mass and relative density and momentum
distributions} The pair center-of-mass and relative density and momentum
distributions Eqs.(\ref{denrelR})-(\ref{denrelk}) are calculated by using
of HO and SW potentials and are given in Figure 6 for the ground state of
$^{16}$O nucleus. They are compared with the results obtained within the
low-order approximation of the Jastrow correlation method (JCM)
\cite{Dim00} where the short-range correlations are explicitely involved.
As it is seen from Figure 6($a'$) the relative distributions
$\rho^{(2)}(\bf {s})$ obtained in GCM and JCM have different behaviour in
the region from 0 to 0.4 fm. This fact shows that the GCM doesn't take into
account short-range but other kind of correlations which are obviously
related with the collective motion of the nucleons. The curves of
${n}^{(2)}(\bf {K})$ calculated for both potentials decrease rapidly down
at K$>$1fm$^{-1}$ while the result obtained within the JCM shows a
high-momentum tail.

\section{Conclusion}
In the present work the characteristics of the ground and the first excited
monopole states of the three nuclei $^{4}$He, $^{16}$O and $^{40}$Ca are
studied within the GCM using different construction potentials. Though the
results are sensitive to the type of the construction potentials used, some
general conclusions can be summarized as follows

(i) There is an increase of the rms radius in the first excited monopole
state as a consequence of the increase of the density distribution for
large $r$.

(ii) There is not an essential difference between the behaviour of the
nucleon momentum distribution in the ground and in the first excited
monopole state.

(iii) The use of the natural orbital representation makes it possible to
study the effective single-particle picture not only of the ground, but
also of the first excited monopole state. It is established that the
depletion of the Fermi sea increases considerably in the first excited
monopole state. This is strongly expressed in the case of $^{4}$He. The
calculated depletion decreases with the increase of the mass number.

(iv) The study of the natural orbitals in the ground and the first excited
monopole state shows that there are not substantial changes in their forms
for a given $nl$-state when they are calculated for the ground and the
first excited monopole state. The rms radii of the hole-state natural
orbitals are larger in the case of the first excited monopole state in
comparison with those for the ground state.

(v) The results on the one- and two-body density and momentum
distributions, occupation probabilities and natural orbitals obtained
within the GCM using various construction potentials show that the
nucleon-nucleon correlations accounted in the approach are different from
the short-range ones but are rather related to the collective motion of the
nucleons. It turns out that the latter are also important in calculations
of one- and two-body overlap functions which are necessary in the
calculations of the cross-sections of one- and two-nucleon removal
reactions. The work on the applications of the GCM overlap functions for
the description of cross-section of $(p,d)$, $(e,e'p)$ and $(\gamma,p)$
reactions is in progress.

\section{Acknowledgments}

\hspace*{1.5em}This work was partly supported by the Bulgarian National
Science Foundation under the Contract No.$\Phi$-809.

\hspace*{1.5em}

\newpage
\begin{center}
{\bf{\Large{Figure Captions}}}
\end{center}
\begin{description}
\item[Figure 1.] One-body distribution function (Eq.(\ref{pro})) $g(r)$ of $^{4}$He
calculated within the GCM using harmonic-oscillator (a), square-well (b)
and Woods-Saxon (c) potentials and within the TIC method \cite{Bis93}. The
normalization is: $\int{{g}_{i}(r)dr}=1
\;\; (i=0,1).$
\item[Figure 2.] Nucleon density distributions of the ground (${\rho}_{0}$) and
the first excited monopole (${\rho}_{1}$) states of $^{16}$O calculated
within the GCM using harmonic-oscillator (a), square-well (b) and
Woods-Saxon (c) potentials. The normalization is:
$\int{{\rho}_{i}({\bf{r}})d{\bf{r}}}=16
\;\;(i=0,1)$.
\item[Figure 3.] The same as in Figure 2 but for $^{40}$Ca.
\item[Figure 4.] Nucleon momentum distributions of the ground (${n}_{0}$) and
the first excited monopole (${n}_{1}$) states of $^{16}$O calculated within
the GCM using harmonic-oscillator (a), square-well (b) and Woods-Saxon (c)
potentials. The normalization is:
$\int{{n}_{i}(k){k}^{2}dk}=16\;\;(i=0,1).$
\item[Figure 5.] Natural orbitals for the 1s-hole (a, a$'$, a$''$), 1p-hole
(b, b$'$, b$''$) and 2s-particle (c, c$'$, c$''$) states of the ground
($i=0$) and of the first excited monopole ($i=1$) states of $^{16}$O in
coordinate space calculated within the GCM using harmonic-oscillator,
square-well and Woods-Saxon potentials, respectively.
\item[Figure 6.] The pair center-of-mass $\rho ^{(2)}({\bf R)}$ (a) and relative $\rho ^{(2)}({\bf s)}$ (a$'$) density distributions and
the pair center-of-mass $n^{(2)}({\bf K)}$ ($b$) and relative $n^{(2)}({\bf
k)}$ (b$'$) momentum distributions of the ground state of $^{16}$O
calculated within the GCM using harmonic-oscillator and square-well
potentials. The results obtained within the Jastrow correlation method
\cite{Dim00} are also given. The normalization is: $\int{{n}^{(2)}({\bf
K)}d{\bf K}}=1$, $\int{{n}^{(2)}({\bf k)}d{\bf k}}=1$, $\int{\rho
^{(2)}({\bf R)}d{\bf R}}=A*(A-1)/2$, $\int{\rho ^{(2)}({\bf s)}d{\bf
s}}=A*(A-1)/2$

\end{description}

\newpage
\begin{table}[ht]
\caption{Energies (in MeV) of the ground and the first excited monopole states in
GCM with square-well, harmonic-oscillator and Woods-Saxon construction potentials
(without the Coulomb energy).\label{t1}}
\begin{center}
\begin{tabular}{cccccc}
 Nuclei & Energies & GCM & GCM & GCM & GCM \\
  &                  &  HO    &  SW    &  WS1    &  WS2  \\
\tableline
$^{4}$He&$E_{0}$ & -29.51 & -37.10 & -26.31 & -28.03 \\
        &$E_{1}$ & -8.06  & -9.87  & -5.38  & -8.29 \\
\hline
$^{16}$O&$E_{0}$ & -139.96 & -144.80 & -137.99 & -139.58 \\
        &$E_{1}$ & -115.01 & -111.66 & -114.09 & -112.50 \\
\hline
$^{40}$Ca&$E_{0}$ & -410.07 & -404.23& -409.23 & -411.40 \\
         &$E_{1}$ & -387.58 & -370.72& -386.13 & -384.92 \\
\end{tabular}
\end{center}
\end{table}

\begin{table}[ht]
\caption{The excitation energies of $0^{+}$ breathing states (in MeV).\label{t2}}
\begin{center}
\begin{tabular}{lcccc}
        &         & $^{4}$He& $^{16}$O& $^{40}$Ca\\
\hline
GCM(HO) & \ \ \ \ & 21.45& 24.95 & 22.49\\
GCM(SW) &         & 27.23& 33.14 & 33.51\\
GCM(WS1)&         & 20.93& 23.90 & 23.10\\
GCM(WS2)&         & 19.74& 27.08 & 26.48\\
Ref. \cite{AnH93}&&      & 23.20 & 27.30\\
Ref. \cite{Bri63}&&      &24     &      \\
Ref. \cite{Pan70}&&      &       &18.3  \\
Ref. \cite{Bru70} ${\eta}$=8&&&30.5&29.5\\
Ref. \cite{Bru70} ${\eta}$=12&&&30.0&26.5\\
Ref. \cite{Bis93}&&24.96 &       &       \\
Ref. \cite{Kam89}&&22.86 &       &       \\
Ref. \cite{Wu 99}&&26.49 & 29.49 &       \\
\hline
     EXP&         &      &  22.9 &   20.0\\
   &&&(T=2) \cite{Ger64} &  (T=0) \cite{Voi85}    \\
\end{tabular}
\end{center}
\end{table}

\begin{table}[ht]
\caption{Rms radii (in fm) of the ground and the first excited
monopole states (first and second row, respectively) in $^{4}$He, $^{16}$O
and $^{40}$Ca obtained with HO, SW, WS1 and WS2 construction
potentials.\label{t3}}
\begin{center}
\begin{tabular}{ccccc}
 Nuclei      & GCM     & GCM     & GCM     & GCM \\
              &  HO    &  SW    &  WS1    &  WS2  \\
\hline
$^{4}$He       &1.89 & 1.77 & 2.28 & 1.85 \\
 & 3.00 & 2.86 & 3.43 & 3.29  \\
\hline
$^{16}$O       & 2.67 & 2.63 & 2.70 & 2.71 \\
 &2.87 & 2.90 & 2.89 & 2.90  \\
\hline
$^{40}$Ca     & 3.37 & 3.40 & 3.38 & 3.39 \\
 & 3.45 & 3.52 & 3.48 & 3.51\\
\end{tabular}
\end{center}
\end{table}

\begin{table} [ht]
\caption{Rms radii (in fm) corresponding to the natural orbitals of the ground and
the first excited monopole state calculated in the GCM with HO, SW, WS1 and
WS2 construction potentials. The Fermi level is denoted by (*).\label{t4}}
\begin{center}
\begin{tabular}{cccccccccc}
\multicolumn{5}{c}{$^{4}$He-ground state}
& \multicolumn{5}{c}{$^{4}$He-first excited monopole state} \\
\hline

state &  HO  &  SW  &  WS1  &  WS2 & state &  HO  &  SW  &  WS1  &  WS2 \\
\hline
   3s & 3.44 & 1.57 &  5.45 & 4.69 &  3s   & 3.40 & 4.92 &  2.68 & 2.91 \\
   2s & 1.78 & 2.77 &  2.62 & 2.84 &  2s   & 2.80 & 2.37 &  3.03 & 4.21 \\
  1s* & 1.88 & 1.76 &  2.26 & 1.84 &   1s* & 3.03 & 2.91 &  3.44 & 3.11 \\
\hline
 \multicolumn{5}{c}{$^{16}$O-ground state} &
\multicolumn{5}{c}{$^{16}$O-first excited monopole state} \\
\hline
state &  HO  &  SW  &  WS1  &  WS2 & state &  HO  &  SW  &  WS1  &  WS2 \\
\hline
   2p & 3.32 & 3.65 &  3.82 & 4.89 &  2p & 3.79 & 3.57 & 4.11 & 3.98 \\
   2s & 2.97 & 3.59 &  2.68 & 4.44 &  2s & 3.36 & 3.55 & 3.82 & 3.29 \\
  1p* & 2.81 & 2.70 &  2.87 & 2.85 &  1p* & 2.97 & 2.92 & 2.97 & 2.96  \\
   1s & 2.18 & 2.35 &  2.13 & 2.21 &  1s & 2.29 & 2.53 & 2.19 &  2.30\\
\hline
\multicolumn{5}{c}{$^{40}$Ca-ground state}
& \multicolumn{5 }{c}{$^{40}$Ca-first excited monopole state} \\
\hline
state &  HO  &  SW  &  WS1  &  WS2 & state &  HO  &  SW  &  WS1  &  WS2 \\
\hline
   3s & 4.22 & 4.48 &4.92 &6.09 & 3s & 4.57 & 4.41 &5.03 &3.68 \\
   2d & 4.28 & 4.55 &4.88 &4.63 & 2d  & 4.58 & 4.51&5.00&3.66 \\
   2p & 3.88 & 4.49 &4.28 &3.08& 2p & 4.14 & 4.46&4.86&3.73 \\
   2s* & 3.62 &  3.56 &3.98 &3.85& 2s* & 3.71 & 3.65&4.02&3.82 \\
   1d & 3.64 & 3.59 &3.63 &3.63&   1d & 3.70 & 3.67 &3.71&3.73\\
   1p & 3.08 & 3.32 &3.06 &3.14&   1p & 3.13 & 3.39 &3.09&3.27\\
   1s & 2.42 & 2.25 &2.04 &2.20&   1s & 2.40 & 2.29 &2.13&2.46\\
\end{tabular}
\end{center}
\end{table}

\begin{table} [ht]
\caption{Occupation numbers and total depletion ${\cal D}$ (Eq.(28)) of the ground and the first
excited monopole state calculated in the GCM with HO, SW, WS1 and WS2
construction potentials. The Fermi level is denoted by (*).\label{t5}}
\begin{center}
\begin{tabular}{cccccccccc}
\multicolumn{5}{c}{$^{4}$He-ground state}
& \multicolumn{5}{c}{$^{4}$He-first excited monopole state} \\
\hline
state &  HO  &  SW  &  WS1  &  WS2 & state &  HO  &  SW  &  WS1  &  WS2 \\
\hline
       3s & $<10^{-3}$ & 0.002 &$<10^{-3}$    &$<10^{-3}$ &
            3s & 0.003 & 0.009 &$<10^{-3}$ &$<10^{-3}$ \\
       2s &      0.005 & 0.014 &0.024&0.006&      2s & 0.149 & 0.152&0.045&0.142 \\
      1s* &      0.995 & 0.984 &0.975&0.993&    1s* & 0.843 & 0.886&0.955&0.8576  \\
\hline
${\cal D}$ & 0.5\% & 1.6\% &2.5\%&0.7\%& ${\cal D}$ & 15.7\%&
 11.4\%&4.5\%&14.2\%\\ \hline
\multicolumn{5}{c}{$^{16}$O-ground state}
& \multicolumn{5}{c}{$^{16}$O-first excited monopole state} \\
\hline
state &  HO  &  SW  &  WS1  &  WS2 & state &  HO  &  SW  &  WS1  &  WS2 \\
\hline
       2p & $<10^{-3}$ & 0.013 & $<10^{-3}$&0.003
       &       2p & 0.067 & 0.095&0.08&0.07 \\
       2s & $<10^{-3}$ & 0.008 & $<10^{-3}$&0.003&
               2s & 0.040 & 0.057&0.02&0.07 \\
      1p* &      0.999 & 0.986 &1.000 &0.999&
            1p* & 0.930 & 0.903&0.933&0.926 \\
       1s &      1.000 & 0.992 &1.000 & 0.999&
              1s & 0.956 & 0.942&0.990&0.955 \\
\hline
${\cal D}$ & $< 0.1$\% & 1.2\% & $< 0.1$\%& $< 0.1$\% & ${\cal D}$ & 6.4\%
& 8.7\%&5.3\%&6.7\% \\
\hline
\multicolumn{5}{c}{$^{40}$Ca-ground state}
& \multicolumn{5}{c}{$^{40}$Ca-first excited monopole state} \\
\hline
state &  HO  &  SW  &  WS1  &  WS2 & state &  HO  &  SW  &  WS1  &  WS2 \\
\hline
       3s & $<10^{-3}$ & 0.014&$<10^{-3}$&$<10^{-3}$ &
               3s & 0.041 & 0.076 &0.038&0.024\\
       2d & $<10^{-3}$ & 0.010 &$<10^{-3}$&$<10^{-3}$&
               2d & 0.029 & 0.055 &0.036&0.024\\
       2p & $<10^{-3}$ & 0.007 &$<10^{-3}$&$<10^{-3}$&
               2p & 0.021 & 0.039 &0.011&0.026\\
      2s* &      0.999 & 0.986 &1.000&0.993&
                   2s* & 0.952 & 0.923&0.960&0.969 \\
       1d &      0.999 & 0.990 &1.000 &0.993&
              1d & 0.964 & 0.945 &0.963&0.970\\
       1p &      0.999 & 0.993 &1.000&0.993&
              1p & 0.972 & 0.961 &0.988&0.968\\
       1s &      1.000 & 1.000 &1.000&0.994&
               1s & 0.993 & 1.000&0.998&0.994 \\
\hline
${\cal D}$ & $< 0.1$\% & 0.9\% &$< 0.1$\%&0.6\%& ${\cal D}$ & 3.2\% &
 4.7\%&2.6\%&2.8\% \\
\end{tabular}
\end{center}
\end{table}

\begin{thebibliography}{99}
\bibitem{AnH88} A.N. Antonov, P.E. Hodgson, I.Zh. Petkov,
{\it Nucleon Momentum and Density Distributions in Nuclei} (Clarendon
Press, Oxford 1988)
\bibitem{AnH93} A.N. Antonov, P.E. Hodgson, I.Zh. Petkov,
{\it Nucleon Correlations in Nuclei} (Springer-Verlag,
Berlin-Heidelberg-New York, 1993)
\bibitem{Boh80} O. Bohigas, S. Stringari, Phys. Lett. {\bf 95B}, 9 (1980)
\bibitem{Jam85} M. Jaminon, C. Mahaux, H. Ng\^{o}, Phys. Lett. B {\bf 158}, 103 (1985)
\bibitem{Fro87} B. Frois, C. Papanicolas, Ann. Rev. Nucl. Part. Sci. {\bf 37}, 133
(1987)
\bibitem{Lap91} L. Lapik\'{a}s, in {\it Proceedings of 6th International Conference
on Nuclear Reaction Mechanisms, Varena, 1991}, edited by E. Gadioli
(Ricerca Scintifica ed Educazione Permanente, Milano, 1991), p. 610
\bibitem{Lap92} L. Lapik\'{a}s, EMIN 92-06, 1992 (Amsterdam)
\bibitem{Hil53} D.L. Hill, J.A. Wheeler, Phys. Rev. {\bf 89}, 1102 (1953)
\bibitem{Gri57} J.J. Griffin, J.A. Wheeler, Phys. Rev. {\bf 108}, 311 (1957)
\bibitem{Sto94} M.V. Stoitsov, P. Ring, M.M. Sharma, Phys. Rev. C {\bf
50}, 1445 (1994)
\bibitem{Vre96} D. Vretenar, G.A. Lalazissis, R. Behnsch, W. P\"{o}schl,
P. Ring, nucl-th/9612042, 1996 (Los Alamos)
\bibitem{Flo75} H. Flocard, D. Vautherin, Phys. Lett. B {\bf 55}, 259 (1975)
\bibitem{Flo76} H. Flocard, D. Vautherin, Nucl. Phys. A {\bf 264}, 197 (1976)
\bibitem{AnC86} A.N. Antonov, Chr.V. Christov, I.Zh. Petkov, Nuovo Cimento A
{\bf 91}, 119 (1986)
\bibitem{AnB88} A.N. Antonov, I.S. Bonev, Chr.V. Christov, I.Zh. Petkov,
Nuovo Cimento A {\bf 100}, 779 (1988)
\bibitem{AnB90} A.N. Antonov, I.S. Bonev, Chr.V. Christov, E.N. Nikolov,
I.Zh. Petkov, Nuovo Cimento A {\bf 103}, 1287 (1990)
\bibitem{AnK94} A.N. Antonov, D.N. Kadrev, P.E. Hodgson, Phys. Rev. C {\bf 50},
164 (1994)
\bibitem{AnB91} A.N. Antonov, I.S. Bonev, I.Zh. Petkov, JINR Rapid
Communications, No 1[40]-90, 35 (1990); Bulg. J. Phys. {\bf 18}, 169 (1991)
\bibitem{AnS94} A.N. Antonov, M.V. Stoitsov, L.P. Marinova, M.E. Grypeos,
G.A. Lalazissis, K.N. Ypsilantis, Phys. Rev. C {\bf 50}, 1936 (1994)
\bibitem{You81} D.H. Youngblood, P. Bogucki, J.D. Bronson, U. Gard, Y.W.
Lui, C.M. Rosza, Phys. Rev. C {\bf 23}, 1997 (1981)
\bibitem{Spe81} J. Speth, A. van der Woude, Rep. Progr. Phys {\bf 44}, 719 (1981)
\bibitem{Voi85} N.A. Voinova-Eliseeva, I.A. Mitropolsky, LINP 1104, 1985
(Leningrad)
\bibitem{Lui85} Y.W. Lui, J.D. Bronson, D.H. Youngblood, Y. Toba, U. Garg,
Phys. Rev. C {\bf 31}, 1643 (1985)
\bibitem{LuB86} H.Y. Lu, S. Brandenburg, R. de Leo, M.N. Harakeh,
T.D. Poelhekken, A. van der Woude, Phys. Rev. C {\bf 33}, 1166 (1986)
\bibitem{Bra87} S. Brandenburg, W.T.A. Borghols, A.G. Drentje, L.P. Ekstr\"{o}m
, M.N. Harakeh, A. van der Woude, A. H{\aa}kansson, L. Nilsson, N. Olsson,
M. Pignanelli, R. de Leo, Nucl. Phys. A {\bf 466}, 29 (1987)
\bibitem{Sha88} M.M. Sharma, W.T.A. Borghols, S. Brandenburg, S. Crona, A. van
der Woude, M.N. Harakeh, Phys. Rev. C {\bf 38}, 2562 (1988)
\bibitem{Bor89} W.T.A. Borghols, S. Brandenburg, J.H. Meier, J.M. Schippers,
M.M. Sharma, A. van der Woude, M.N. Harakeh, A. Lindholm, L. Nilsson, S.
Orona, A. H{\aa}kansson, L.P. Ekstr\"{o}m, N. Olsson, R. de Leo, Nucl.
Phys. A {\bf 504}, 231 (1989)
\bibitem{AnN91} A.N. Antonov, V.A. Nikolaev, I.Zh. Petkov, Bulg. J. Phys
{\bf 18}, 107 (1991)
\bibitem{Bru70} K.A. Brueckner, M.J. Giannoni, R.J. Lombard, Phys. Lett. B
{\bf 31}, 97 (1970)
\bibitem{Pan70} V.R. Pandharipande, Phys. Lett. B {\bf 31}, 635 (1970)
\bibitem{Tre81} J. Treiner, H. Krivine, O. Bohigas, J. Martorell, Nucl. Phys. A
{\bf 371}, 253 (1981)
\bibitem{VoE85} N.A. Voinova-Eliseeva, I.A. Mitropolsky, LINP 1095, 1985
(Leningrad)
\bibitem{Sha89} M.M. Sharma, W. Stocker, P. Gleissl, M. Brack, Nucl. Phys. A
{\bf 504}, 337 (1989)
\bibitem{Bis93} R.F. Bishop, E. Buendia, M.F. Flynn, R. Guardiola, J. Phys. G
{\bf 19}, 1163 (1993)
\bibitem{Wu 97} J.S. Wu, K.C. Wong, M.R. Strayer, M. Baranger, Phys. Rev. C
{\bf 56}, 857 (1997)
\bibitem{Col92} G. Col\`{o}, N. Van Giai, P. Bortignon, A. Bracco, R.
Broglia, Fizika {\bf 22}, 267 (1990); Phys. Lett. B {\bf 276}, 279 (1992)
\bibitem{Wu 99} J.S. Wu, M.R. Strayer, M. Baranger, Phys. Rev. C {\bf 60},
044302 (1999)
\bibitem{Ma 97} Z. Ma, N. Van Giai, H. Toki, M. L'Huillier, Phys. Rev. C
{\bf 55}, 2385 (1997)
\bibitem{Lwd55} P.-O. L\"{o}wdin, Phys. Rev. {\bf 97}, 1474 (1955)
\bibitem{BH 91} H.P. Blok and J.H. Heisenberg, in {\it Computational Nuclear
Physics 1, Nuclear Structure}, Edited by K. Langanke, J.A. Maruhn, S.E.
Koonin (Springer-Verlag, Berlin, 1991), p. 190.
\bibitem{Bri66} D.M. Brink,in {\it Proc. S I F, Course XXXVI}, edited by C. Bloch
(Academic Press, New York, 1966), p. 247
\bibitem{AnC89} A.N. Antonov, Chr.V. Christov, E.N. Nikolov, I.Zh. Petkov,
P.E. Hodgson, Nuovo Cimento A {\bf 102}, 1701 (1989)
\bibitem{Bar82} J. Bartel, P. Quentin, M. Brack, C. Guet,
H.B. H{\aa}kansson, Nucl. Phys. A{\bf 386}, 79 (1982)
\bibitem{Bri63} D.M. Brink, G.F. Nash, Nucl. Phys. {\bf 40}, 608 (1963)
\bibitem{Kam89} H. Kameyama, M. Kamimura, Y. Fukushima, Phys. Rev. C {\bf 40},
974 (1989)
\bibitem{Cso97} A. Cs\'{o}to, G.M. Hale, Phys. Rev. C {\bf 55}, 2366
(1997)
\bibitem{Ger64} J. Cerny, R.H. Pehl, G.T. Garvey, Phys. Lett. {\bf 12}, 234 (1964)
\bibitem{Dim00} S.S. Dimitrova, D.N. Kadrev, A.N. Antonov, M.V. Stoitsov,
nucl-th/9904071; Condensed Matter Theories, v.15, Nova (in print)
\end{thebibliography}
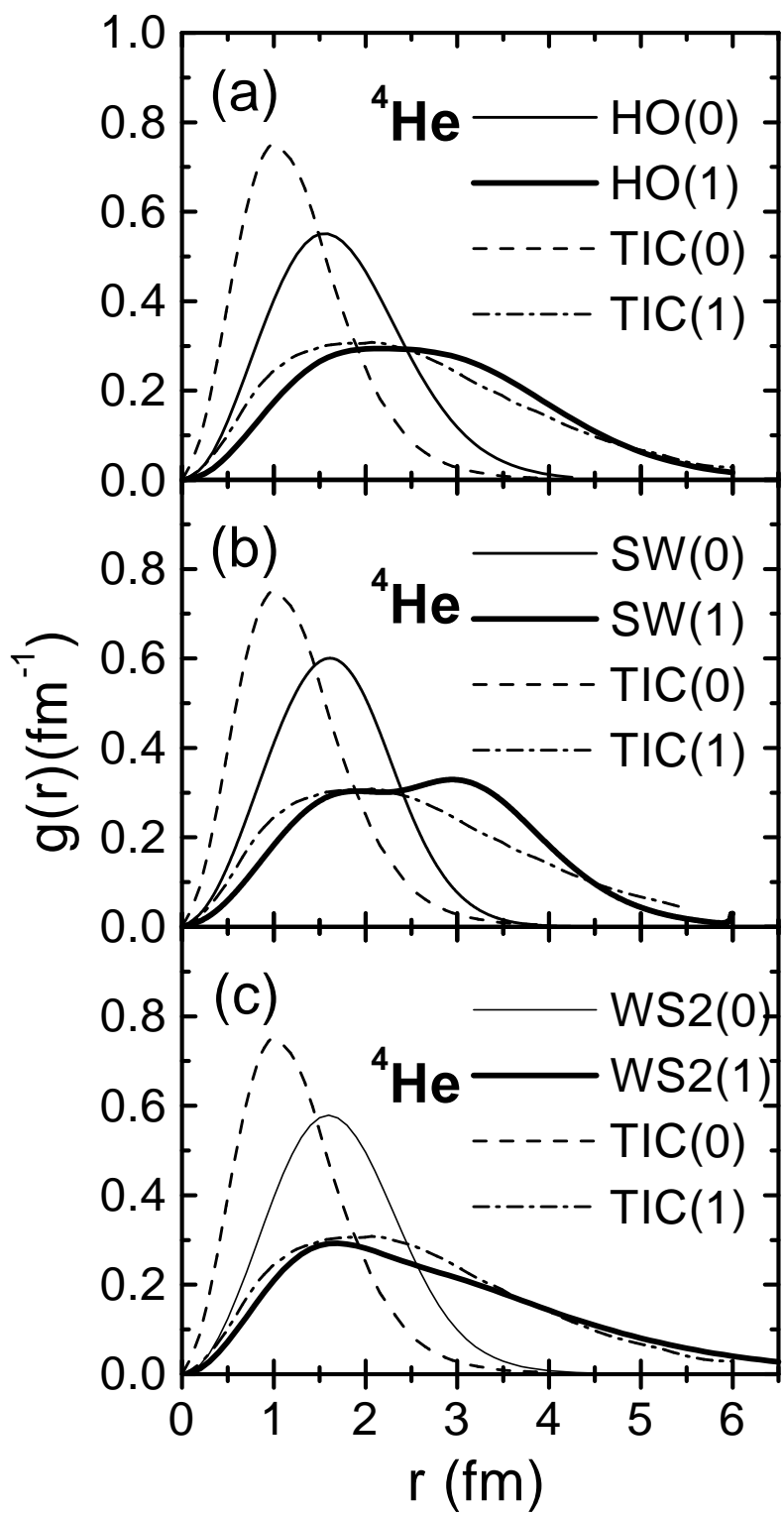
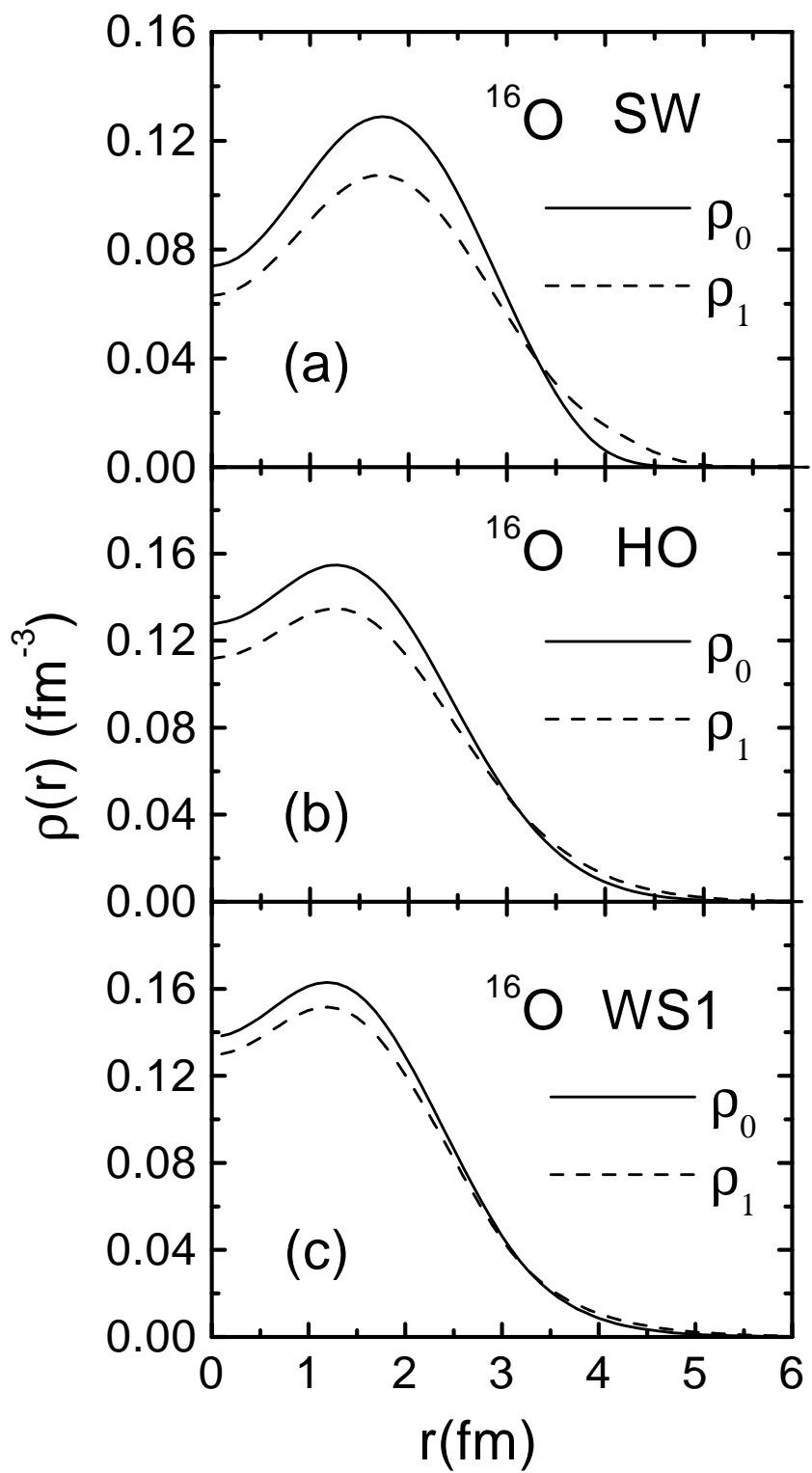
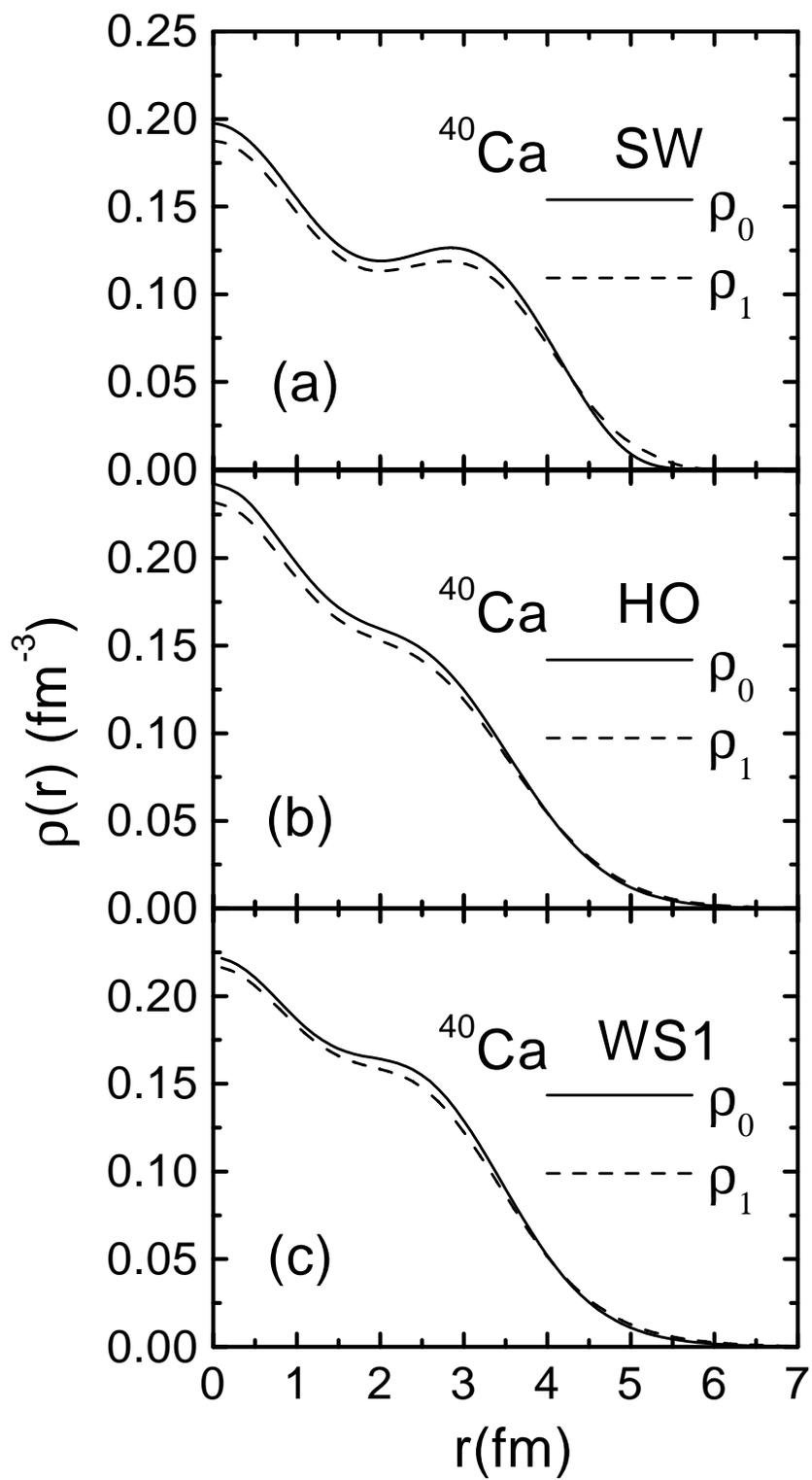
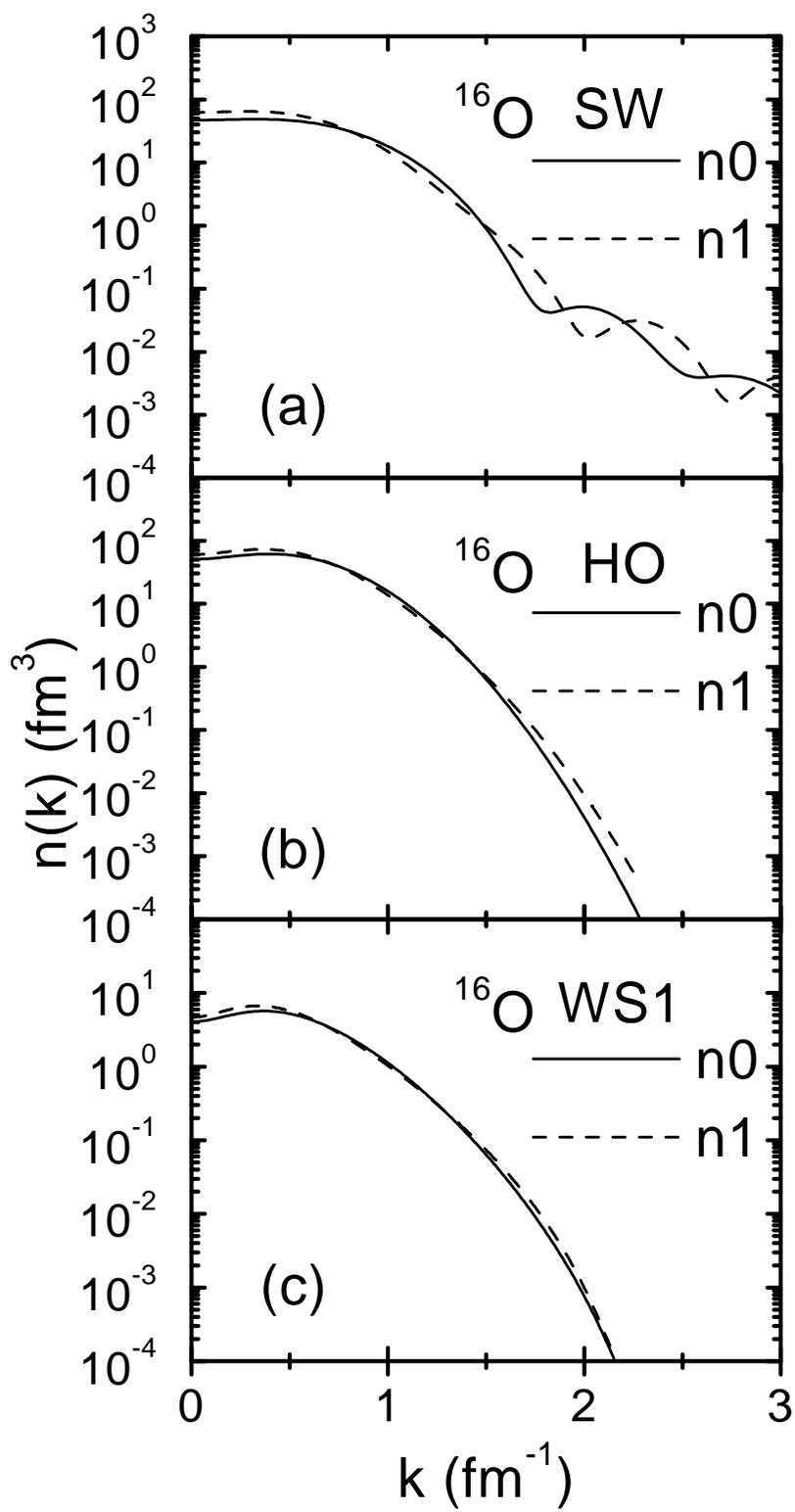
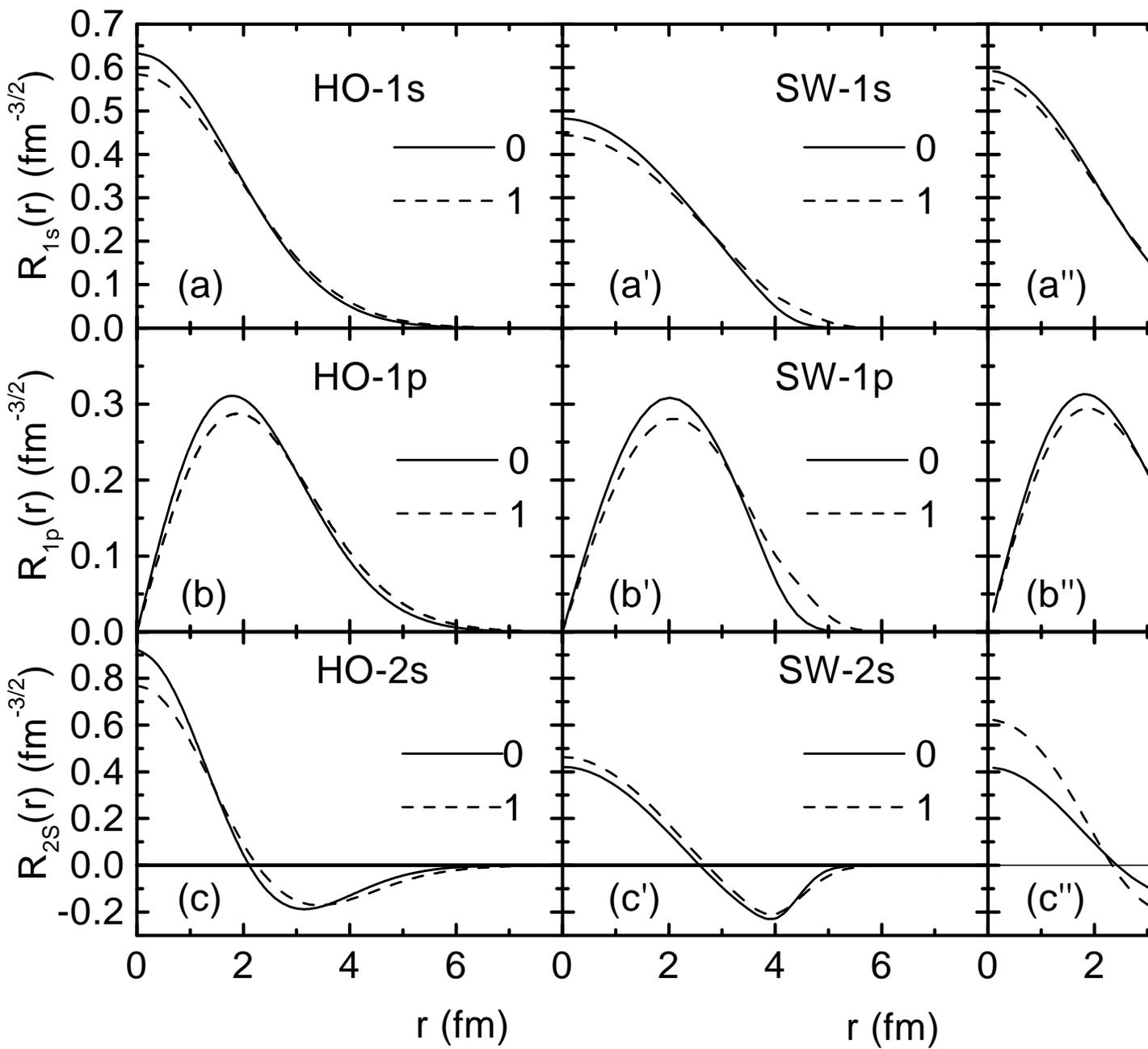
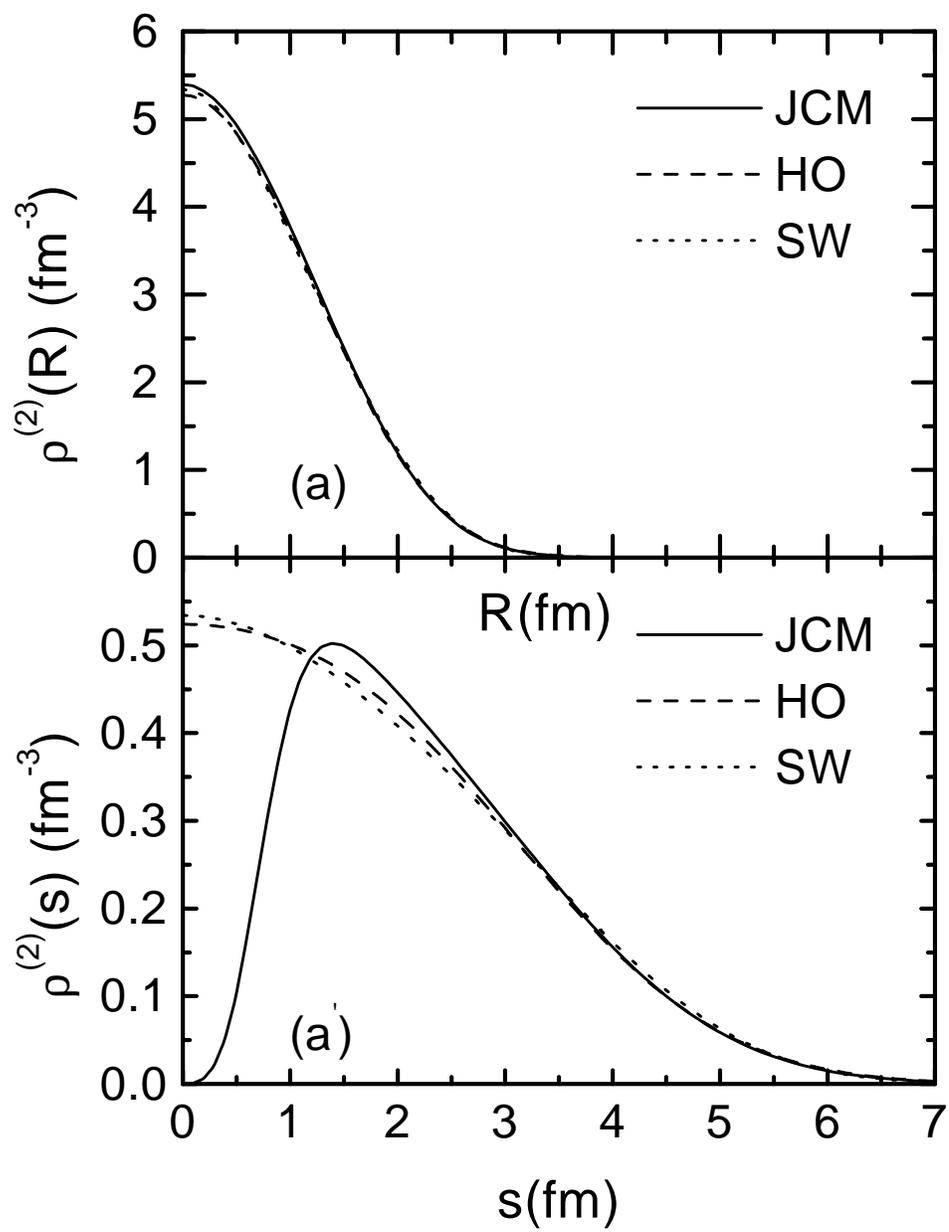
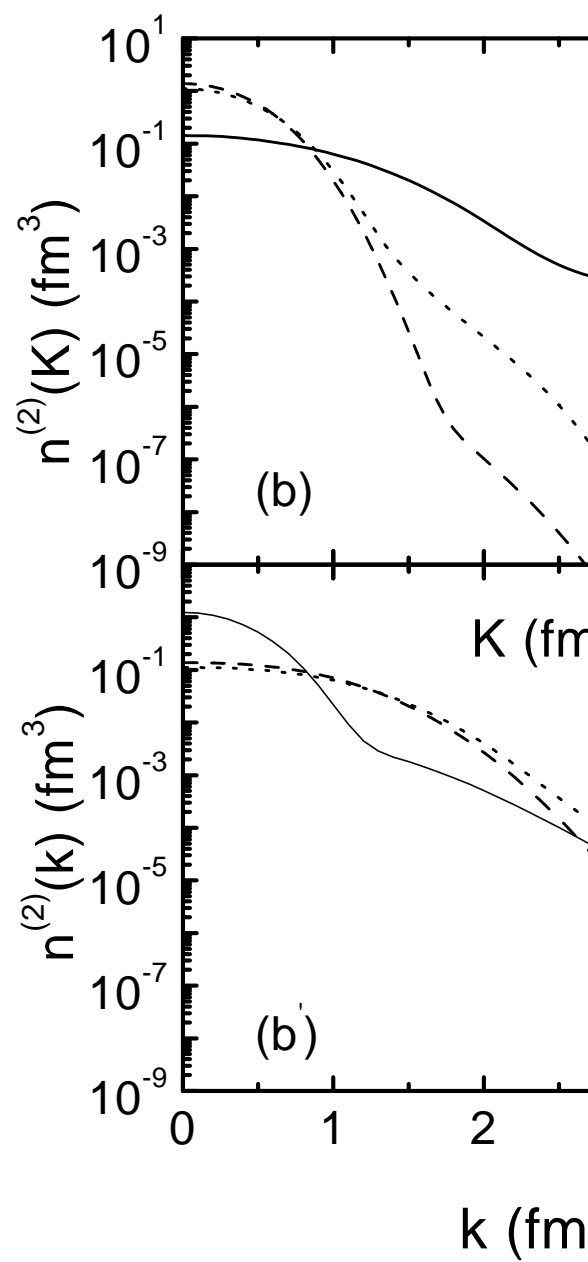
\end{document}